\documentclass[journal]{IEEEtran}
  \usepackage{graphicx}
  \DeclareGraphicsExtensions{.pdf,.jpeg,.png,.eps}
\usepackage[justification=centering]{caption}
\usepackage[cmex10]{amsmath}
\usepackage{algorithmic}
\usepackage{array}
\usepackage{mdwmath}
\usepackage{mdwtab}
\usepackage{tabularx}
\usepackage{booktabs}
\usepackage{multirow}
\usepackage{bigstrut}
\usepackage[switch,columnwise]{lineno} 
\usepackage{cite}
\usepackage{pdfpages} 
\usepackage{color}
\usepackage{soul}

\begin{document}

\title{An Outlook on the Interplay of Machine Learning and Reconfigurable Intelligent Surfaces: An Overview of Opportunities and Limitations}

\author{Lina Mohjazi, Ahmed Zoha, Lina Bariah, Sami Muhaidat, Paschalis C. Sofotasios, Muhammad Ali Imran, and Octavia A. Dobre \\
*This article is published in IEEE Vehicular Technology Magazine, vol. 15, no. 4, Dec. 2020, under title \\ "An Outlook on the Interplay of Artificial Intelligence and Software-Defined Metasurfaces: An Overview of Opportunities and Limitations"

 \thanks{L. Mohjazi, A. Zoha and M. A. Imran are with the James Watt School of Engineering, University of Glasgow, Glasgow, U.K (e-mail: \{Lina.Mohjazi, Ahmed.Zoha, Muhammad.Imran\}@glasgow.ac.uk).}
  \thanks{L. Bariah, S. Muhaidat, and P. C. Sofotasios are with the Center for Cyber-Physical Systems, Khalifa University, Abu Dhabi, UAE (e-mail: \{lina.bariah, muhaidat, p.sofotasios\}@ieee.org).}
\thanks{O. A. Dobre is with the Department of Electrical and Computer Engineering, Memorial University, St. John’s, Canada (e-mail: odobre@mun.ca).}}

\markboth{Accepted in IEEE Vehicular Technology Magazine}%
{Shell \MakeLowercase{\textit{et al.}}: Bare Demo of IEEEtran.cls for Journals}
\maketitle

\begin{abstract}
Recent advances in programmable metasurfaces, also dubbed as reconfigurable intelligent surfaces (RISs),  are envisioned to offer a paradigm shift from uncontrollable to fully tunable and customizable wireless propagation environments, enabling a plethora of new applications and technological trends. Therefore, in view of this cutting edge technological concept,  we first review the architecture and  electromagnetic waves manipulation functionalities of RISs. We then detail some of the recent advancements that have been made towards realizing these programmable functionalities in wireless communication applications. Furthermore, we elaborate on how machine learning (ML) can address various constraints introduced by  real-time deployment of RISs, particularly  in terms of latency, storage, energy efficiency, and computation. A review of the state-of-the-art research on the integration of ML with  RISs is presented, highlighting their potentials as well as challenges. Finally, the paper concludes by offering a look ahead towards unexplored possibilities of ML mechanisms in the context of RISs.
 \end{abstract}

\IEEEpeerreviewmaketitle

\section{Introduction}
\label{intro}
The unprecedented proliferation of connected devices, driven by the emergence of the Internet of Everything (IoE), has created a major challenge for broadband wireless networks, which would require a paradigm shift towards the development of key enabling technologies for the next generation of wireless networks. This explosive growth is coupled with technology revolutions and new societal trends that are expected to shape future breakthrough IoE-enabled services, such as virtual reality, augmented reality,  telemedicine, flying vehicles, holographic telepresence, and connected autonomous artificial intelligence (AI) systems through machine-to-machine communications \cite{zappone2019wireless}.
\par The fifth generation (5G) wireless networks have been identified as the backbone of emerging IoE services, and  prominently support three use cases:  enhanced mobile broadband, ultra reliable and low-latency communications, and massive machine-type communications. These services are rate and data-oriented and heterogeneous in nature, which are defined by a diverse set of key performance indicators. Therefore, enabling them through a single platform while concurrently meeting their stringent requirements in terms of data rate, reliability and latency, is a challenging task \cite{zappone2019wireless}.
\par To address the aforementioned challenges at the physical layer, 5G leveraged the evolution of cutting edge technologies, such as  millimeter wave (mmWave) communications,  ultra-dense networks, and massive multiple-input multiple-output (MIMO) communications. Although these technologies have significantly improved the efficiency of wireless networks, the associated hardware complexity, cost and  increasing energy consumption still pose critical challenges for their practical implementation. In particular, mmWave and  Terahertz (THz) communications are considered two of the most promising technological paradigms in future wireless networks, offering unparalleled data rates and significantly reducing the required device size. However, their present use is limited due to signal degradation at these extremely high communication frequency bands.  Moreover,  wireless links suffer from attenuation incurred by high propagation loss, high penetration loss, multi-path fading, molecular absorption, and Doppler shift \cite{Liaskos1}. With the lack of full control over the propagation and scattering of electromagnetic (EM) waves, the wireless environment remains unaware of the time-variant communication, posing fundamental limitations towards building truly pervasive software-defined wireless networks \cite{Liaskos1}.
\begin{figure*}[!t] 
\centering
\captionsetup{justification=centering}
\includegraphics[width=7.3in]{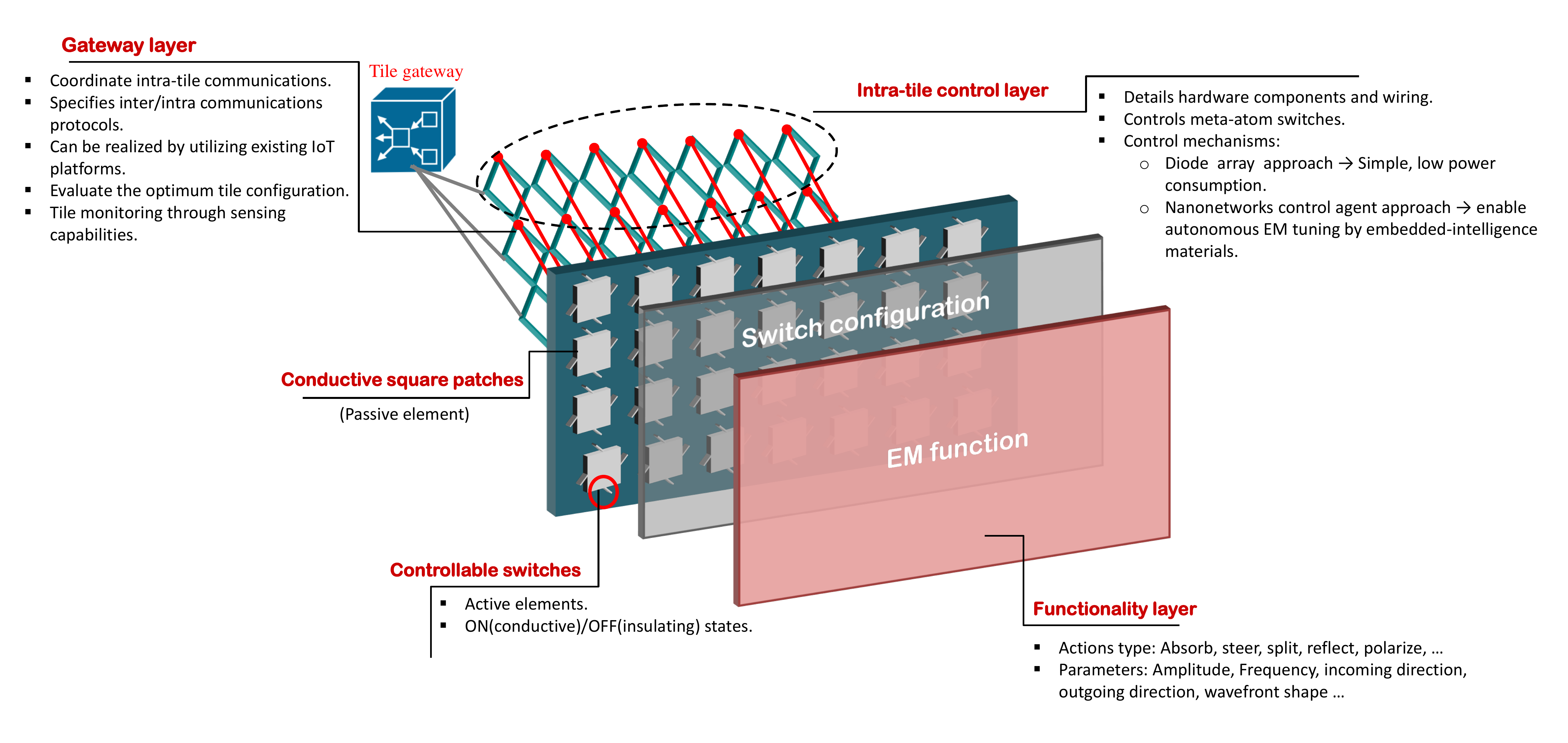}
\caption{Architecture of RISs \cite{Liaskos1}.}
\label{fig:Arch}
\end{figure*}
\par Motivated by the need to develop innovative low-complexity and energy-efficient solutions, the concept of reconfigurable metasurfaces, also known as reconfigurable intelligent surfaces (RISs), has emerged as a revolutionary technology that aims at turning the wireless environment into a software-defined entity \cite{Liaskos1}. 
\par RISs are envisaged to be indispensable in \textit{the sixth generation (6G) wireless systems},   due to their potential in realizing the massive MIMO gains, while attaining a notable reduction in energy consumption \cite{Liaskos1}. The unique design principle of RISs lies in realizing artificial structures with massive antenna arrays, whose interaction with the impinging EM waves can be intentionally controlled through connected passive elements, such as phase shifters, in a way that enhances the performance of wireless systems in terms of coverage, rate, etc., giving rise to the concept of "smart radio environments" (SREs).
\par In light of this, AI tools are envisioned to be intrinsic in RISs to  identify the best operation policy based on data-driven techniques. Specifically, the application of machine learning (ML), as a subfield of AI, is foreseen to play a key role in RISs for a wide variety of applications \cite{Huang}. This stems from its capability to dynamically change the paradigm of data processing through the employment of algorithms that can learn from data and perform functionalities to complete complex tasks efficiently.
\par The main contribution of this paper is a forward looking vision of ML-empowered RISs. Specifically, in Section \ref{architecture}, we identify their operational principle and wireless functionalities, highlighting their potential applications and associated practical challenges. Section \ref{AI SDMs} details the potentials and limitations of several fundamental ML categories in optimizing the performance of RISs-enabled networks. A look ahead towards the implementation of some interesting ML mechanisms in configuring RISs in future deployments is offered in Section \ref{lookahead}. Finally, concluding remarks are presented in Section \ref{conc}. 
\section{Metasurfaces: Architecture, Applications, and  Challenges}\label{architecture}
In this section, we elaborate on the working principle of RISs and present an overview for some of the applications. This is followed by discussing the associated challenges from which we derive our vision as to how ML-based approaches can potentially optimize the design network parameters of RISs. 
\subsection{Metasurfaces Architecture}
Conventional metasurfaces, based on fixed phase shifters, can be found in several applications, e.g., satellite and radar communications. However, their commercialization is limited to a few companies, such as Mast Technologies$^{\copyright}$ and RF Microtech$^{\copyright}$, since once manufactured, they perform a specific functionality and their parameters cannot be reconfigured. 
\begin{table*}
\centering
\caption{Hardware Architecture Aspects of RISs Main Operational Principles.}
\vspace*{-1cm}
\includegraphics[page={1},width=6.9in]{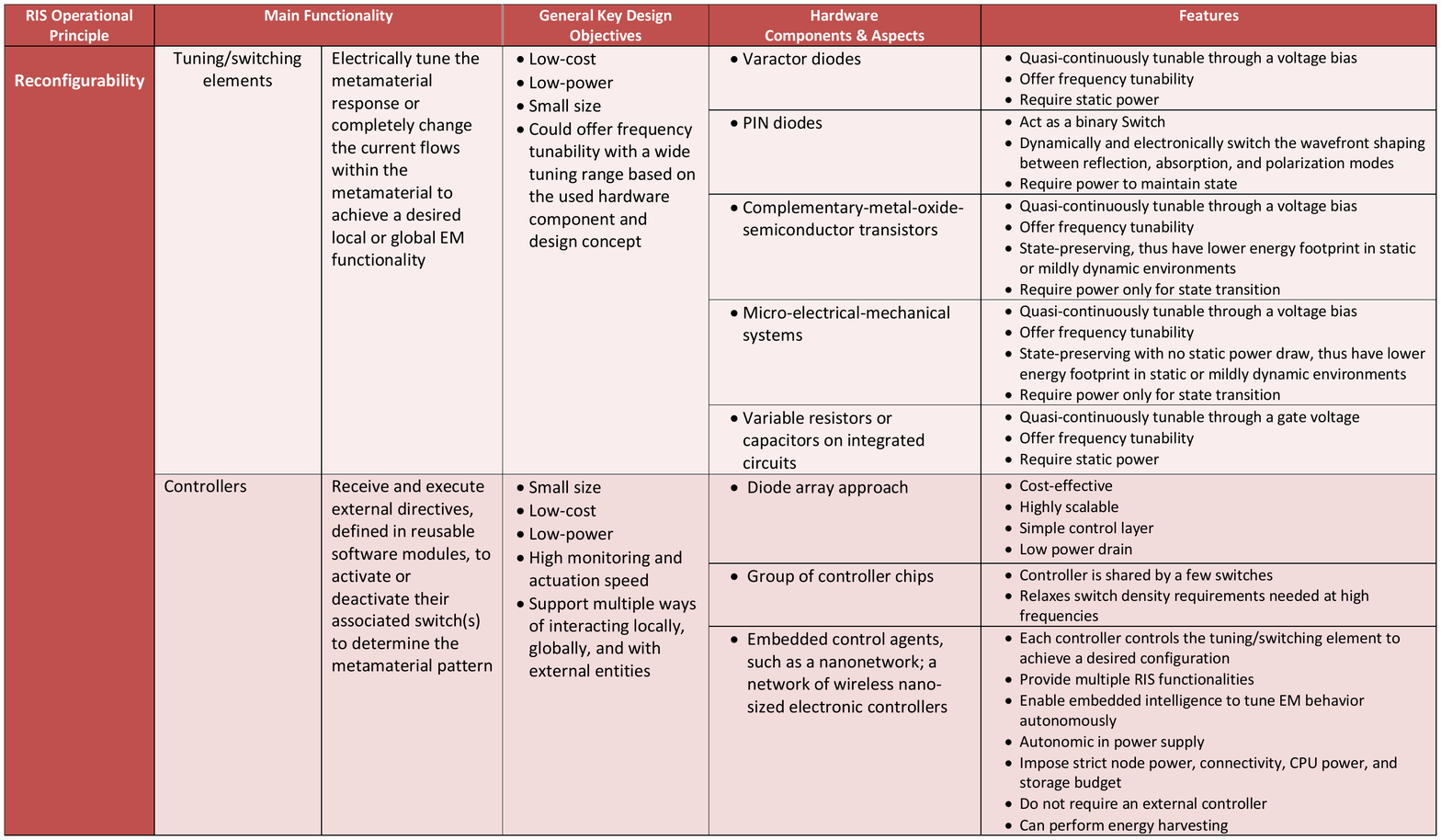}
\vspace*{-0.36cm}
\vspace*{-2.86cm}
\includegraphics[page={2},width=6.9in]{SDM_hardware.pdf}
\vspace*{-0.48cm}
\vspace*{-3cm}
\includegraphics[page={3},width=6.9in]{SDM_hardware.pdf}
\label{hard}
\end{table*}
\par The metasurface architecture, comprising meta-atoms as its building block,  is a sophisticated metallic or dielectric small scattering particle that is periodically repeated over a certain area to create a planar (also called a tile) \cite{Liaskos1,9007666}. Owing to their artificially engineered structures, metasurfaces enable unprecedented capabilities in interacting with the impinging EM waves, such as wave focusing, absorption, imaging, scattering, polarization, to name a few \cite{Liaskos1,9007666}. Thus, when treated macroscopically, the permittivity and permeability of metamaterials are completely defined by the meta-atom structure and can be customized locally into any configuration of choice to shape the impinging EM waves, eliminating the need to generate new radio signals, and reducing the overall network energy consumption.
\par Programmable metasurfaces, also referred to as RISs, have been recently introduced to realize the vision of SRE by leveraging software control methodologies. RISs have been considered for terrestrial wireless communication applications only recently, due to advancements in tunable switching components, which are fitted in the meta-atoms and receive commands from an external programming interface consisting of software-defined functions that alter the meta-atom structure dynamically and, thereby, reconfigure and manipulate the parameters of incident and reflected waves, e.g.,  phase, amplitude, frequency, and polarization, to enable the EM behavior of interest \cite{9007666}. The architecture of RISs is depicted in Fig. \ref{fig:Arch}. The tunable mechanisms of metasurface allow for a variety of functionalities at different frequencies including mmW and THz bands, facilitating massive connectivity,  interference mitigation, and enhanced diversity by introducing an additional degree of freedom. A summary of the hardware architecture, proposed in the open literature \cite{zappone2019wireless, Liaskos1, 9007666} to realize key operational principles of RISs, is provided in Table~\ref{hard}. Although several prototypes are proposed to realize reconfigurable RISs in real time (e.g., \cite{Liaskos1} and the references therein), their full commercialization is still underway.
\par The introduction of RISs \cite{Liaskos1} facilitated turning the wireless environment into a programmable and partially deterministic space, which can be incorporated as a parameter in the network design. Some  \textit{metasurface} functionalities include, but not limited to, beam steering, beam splitting, wave absorption, wave polarization, and phase control. The interested readers are referred to \cite{Wang} for a detailed overview of RIS antenna design principles and EM working topologies.
\subsection{Metasurfaces Applications}
In the following, we discuss some of the potential RISs applications.
\subsubsection{\textbf{Multi-User Communications}}
In \cite{8741198}, metasurfaces were investigated for downlink transmission scenarios to support multiple users through beam splitting and beamforming/beam steering, as illustrated in Fig.~\ref{fig:App}(a), shown on the next page.
\subsubsection{\textbf{Signal Modulation}}
Due to their unique properties, metasurfaces have been proposed as low-cost and low-energy signal modulators \cite{Shan} (see application scenario in Fig.~\ref{fig:App}(a)).
\subsubsection{\textbf{Wireless Power Transfer (WPT)}}
The unique properties of metasurfaces, which include their abilities to steer and concentrate EM waves, enable efficient WPT to provide perpetual energy replenishment, particularly for low-power devices/sensors, as depicted in Fig.~\ref{fig:App}(b) \cite{Liaskos1}. 
\subsubsection{\textbf{Relaying Communications}}
One of the attractive applications of metasurfaces is to operate as a passive reflective relay \cite{ntontin2019reconfigurable}, at a reduced hardware complexity and power consumption, to enhance the QoS of users suffering from detrimental propagation conditions, as shown in Fig.~\ref{fig:App}(c).  
\subsubsection{\textbf{Physical Layer Security}}
\begin{figure*}[!t] 
\centering
\includegraphics[width=7.2in]{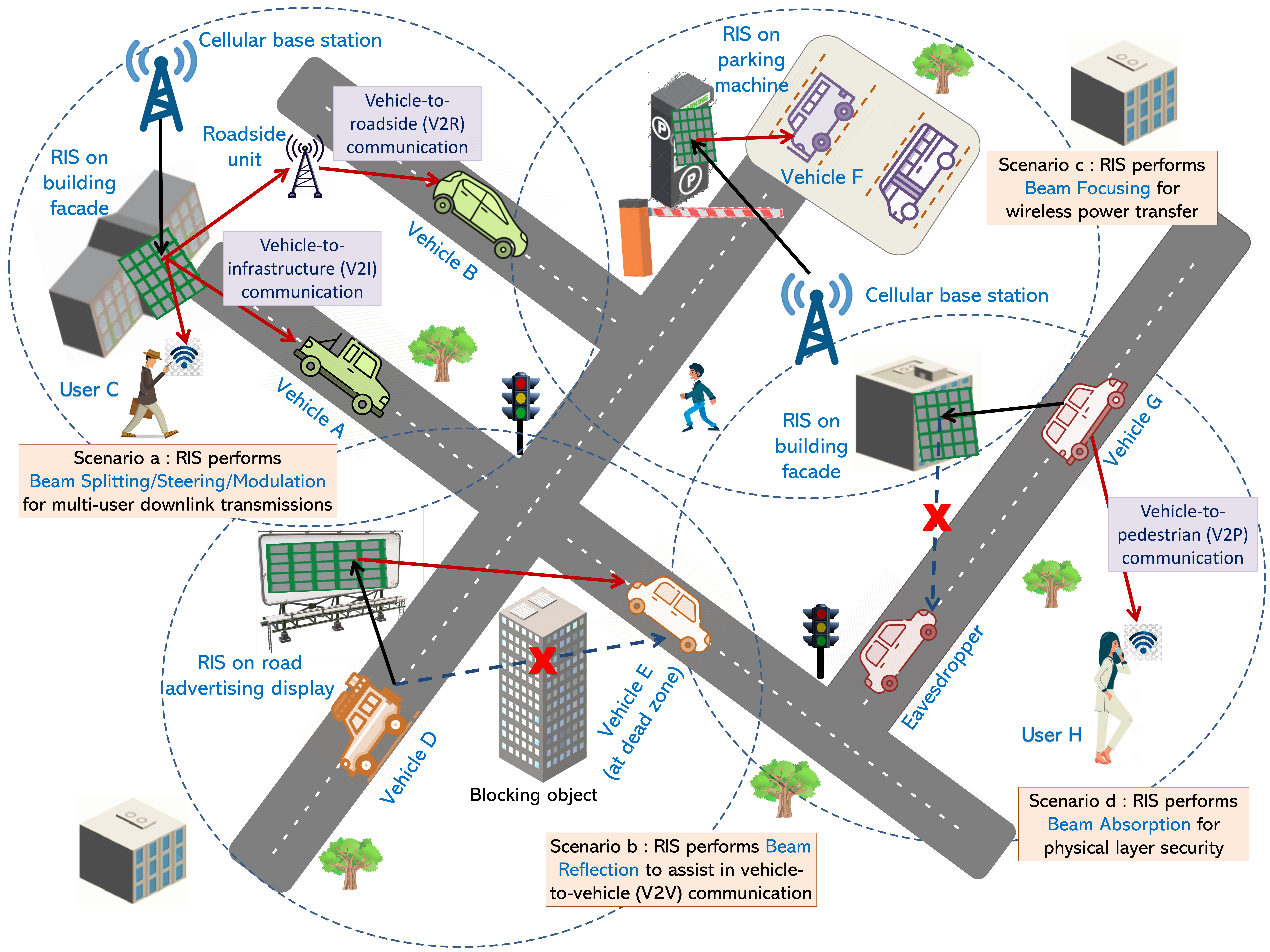}
\caption{Typical RIS applications envisioned for future smart vehicular-to-everything (V2X) environments: (a) Multi-User Communications (b) Relaying Communications (c) Wireless Power Transfer (d) Physical Layer Security.}
\label{fig:App}
\end{figure*}

The design of beamforming for metasurfaces-assisted secrecy communications is examined to improve the secrecy rate of legitimate receivers in the presence of an eavesdropper \cite{MS}. This scenario is illustrated in Fig.~\ref{fig:App}(d).\\
\indent Both RISs' functionalities and applications are envisioned in the context of future smart vehicular-to-everything (V2X) environments, as shown in Fig.~\ref{fig:App}.


%
%
%
\subsection{Key Technical Challenges}
Despite the promising prospects of metasurfaces in 6G, there exist several critical challenges which need to be addressed for realizing the full potentials of this technology. Some of them are described in what follows.

\subsubsection{\textbf{Latency}}
The speed of adaptivity and reconfigurability of the metasurfaces' EM response is a crucial aspect for the successful practical implementation of the aforementioned emerging metasurface-based applications. However, in practice, the number of wireless nodes connected to a metasurface can be significantly high, for example in case of IoEs scenarios. In this case, the number of parameters associated with the system optimization increases. Therefore, the convergence time needed to reach an optimal solution becomes prohibitive.   
\subsubsection{\textbf{Storage \& Energy Efficiency}}
Since the operation of RISs is primarily based on sensing, computing, and information processing to enable SREs, the amount of sensed data and overhead-feedback required to optimize the configuration of the EM response of metasurfaces is excessively high.  This in turn increases the demands for additional resources, not only in terms of computational time and storage, but also in terms of energy and bandwidth. 
\subsubsection{\textbf{Computation}}
Future SREs are also envisaged to be heterogeneous, where different sophisticated technologies supporting diverse requirements with different QoS coexist. The optimization of RIS parameters to coordinate  processes  and  enable uninterrupted connectivity is a non-trivial task. This calls for new computation algorithms to efficiently and dynamically adapt network parameters, such as coding rate, route selection, frequency band, and symbol modulation.
\subsubsection{\textbf{Analytical Models}}
Mathematical models are necessary for initial network planning, resource management, and network control. However, traditional approaches might not be feasible to successfully model the operation of intelligent tunable metasurfaces in future wireless networks. The reason is the exponential increase in system complexity associated with the explosive growth in the number of connected devices. Since metasurfaces are envisioned to be attached to environmental objects, such as walls, facades of buildings, or even as a part of fabrics, their spatial distribution patterns are very complex. This renders existing mathematical models, developed based on assumptions that are inconsistent with the physical response of metasurfaces, to be inapplicable. 
\par In order to realize the vision of limitless and seamless connectivity, ML-enabled solutions can be presented to support self-organization and automation of all RIS functions, including maintenance, management, and operational tasks. Owing to their ability to solve complex problems, ML-based mechanisms will be discussed in the next section, where we highlight some of the potential enabling applications in the area of RISs. 
\section{ML-Enabled Metasurfaces}\label{AI SDMs}
An RIS  requires integration of its metamaterial structure with a network controller in order to realize the desired EM behavior and to support real-time adaptivity of its functionalities \cite{Liaskos1}. Specifically, certain metasurface attributes, e.g., patterns, bias, or impedance, may be determined by activating or deactivating the associated switches on the controllers. Based on that, multiple RIS functionalities may be achieved concurrently and adaptively by enabling reusable software modules, which are able to incorporate efficient learning mechanisms.
\par RISs would necessitate a complex level of coordination to maintain their desired global behavior while ensuring scalability and energy and overhead reduction. Accordingly, ML techniques appear to be promising for enabling intelligence in RIS interfaces, given that the number of their embedded sensors and controllers is anticipated to increase rapidly \cite{Huang}. Table~\ref{hard} also suggests that network resource management, taking into consideration memory space, computing power, bandwidth, traffic demands, network requirements, etc., is an extremely challenging task \cite{zappone2019wireless}.
\par ML techniques have been extensively investigated in wireless networks \cite{simeone2018very}, since they can offer practical and elegant solutions to frequent problems arising in wireless communications, such as classification, optimization, estimation, and detection. Nonetheless, to the best of our knowledge, a comprehensive survey of the state-of-the-art on key ML enablers for RIS communication environments does not exist in literature.  To address this, in what follows we shed light on classical ML paradigms and major learning frameworks, detailing their potentials and limitations in the context of RISs.
\par Learning algorithms available in literature are classified
into three main categories: supervised, unsupervised, and reinforcement. A separate subdivision of ML that is extensively discussed in literature is deep learning (DL). DL algorithms emulate human brain by using complex multi-layered neural networks, and extracting high-level, complex abstractions as data representations through a hierarchical learning process. The learning process can be any of the learning principles mentioned above.
\subsection{Classical Machine Learning Paradigm and RISs: Can They Handshake?}
In recent years, classical ML algorithms, such as decision-tree, Gaussian mixture models, k-nearest neighbor, support vector machine models as well as rule-based and inductive logical programming models, have proven to be advantageous in assisting and enhancing the operation and design of wireless communication systems\cite{simeone2018very}. The two most common categories of ML algorithms, namely supervised and unsupervised learning algorithms, are briefly described as follows:
\begin{itemize}
    \item \textbf{Supervised learning}: In this approach, both input (training set) and output (labels) data are necessary for  algorithms to learn and an explicit input-output mapping is required~\cite{simeone2018very}.\\
    \item \textbf{Unsupervised learning}: In this method,  algorithms must be able to extrapolate the statistical structure of the input or any other given information to proceed with a specific task.   
\end{itemize}  
\par The existing literature discusses plethora of classical ML applications across different layers of communication networks ranging from channel estimation, signal focusing, user-cell association, network optimization, resource scheduling, and intelligent beamforming~\cite{simeone2018very}. 
\par \textbf{\textit{Challenges:}}
Despite their efficacy for existing systems, classical ML models are not adequate for RISs due to a number of factors: (i) their dependence on availability of large amounts of data to achieve low generalization error; (ii) inability to meet the stringent requirements of ultra-low latency communication due to large processing delays; (iii) absence of closed-loop optimization functionality to interact and incorporate response from the dynamic environment surrounding RISs. 
\subsection{Closed-loop Learning for RISs}\label{cll}
The advancement in learning strategies, such as reinforcement learning (RL) algorithms are more suited to match the operational requirements in a highly dynamic SRE. In RL, a feedback loop exists between the environment and the algorithm, allowing it to adaptively converge to an ideal behavior leveraging the feedback received from the environment. These closed-loop/self-learning algorithms
autonomously observe the change in sensed data over time without supervision and accordingly, their only source of knowledge is derived from their environment following a blind action execution strategy, aiming to maximize the target reward (performance). They are sometimes referred to as self-supervised learning systems. Traditionally, RL approaches have been employed to address various challenges in conventional communication systems, including power control, antenna tilt optimization, data offloading, and adaptive modulation. Within the ambit of SRE applications, a self-supervised RL algorithm has been studied that enables a feedback loop between the software controller and the metasurface response to the radio waves. The controller leverages the self-learning strategy of the RL algorithm and jointly optimizes the sensed data and wave manipulations applied by RISs~\cite{liaskos2019interpretable}. For example, steering of multipath reflection to an intended destination could only be envisaged if RISs have the capability to operate in a closed-loop fashion; meaning they interact with the environment, take actions, and subsequently, incorporate the resulting feedback for optimized decision making. However, this comes at the cost of adding a feedback circuit that interconnects the sensors and the programming unit, such as an FPGA, whose features are presented in Table~\ref{hard}. Likewise, metasurface-based modulation techniques show promising results in collecting essential contextual information from the environment and encoding it onto the reflections of other signals as a feedback to optimize the operation of RISs~\cite{Shan}. 
\par \textbf{\textit{Challenges:}}
RL algorithms suffer from the bottleneck of longer convergence time due to exploitation-exploration, making them nonviable for realizing ultra-low latency driven RIS wireless functionalities. 
\begin{figure*}[!t] 
\centering
\includegraphics[width=6.5in]{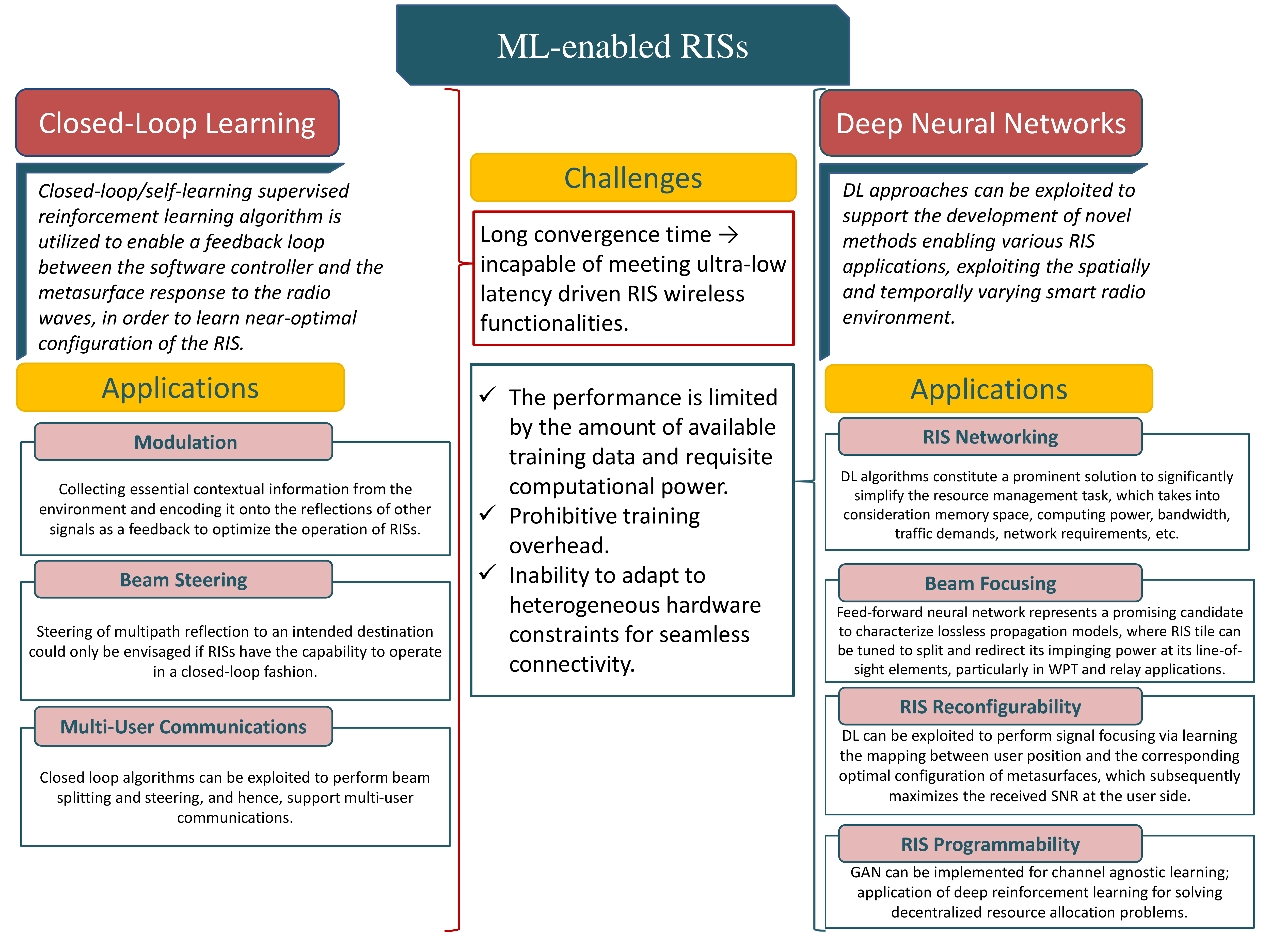}
\caption{Potential ML enablers for RISs.}
\label{fig:AI}
\end{figure*}
\subsection{Deep Neural Networks: A Way Forward?}\label{dnn}
Compared to the aforementioned learning approaches, DL algorithms have garnered significant attraction in the last couple of years, especially in the domain of wireless communications. The ability of the DL approaches to learn and represent a correlational structure in the available data with or without prior domain knowledge clearly provides a decisive advantage over traditional approaches. This is achieved via a neural network architecture with multiple hidden layers capable of discovering latent structures within labeled, unstructured and unlabeled data by proceeding it in a supervised, reinforcement, unsupervised, or hybrid fashion. The application of DL-based solutions to wireless communications has shown a great promise across all layers, from physical to network development and planning~\cite{Shan,zhang2019deep}. At the physical layer, DL-assisted solutions have been deployed to simplify various tasks such as channel estimation and decoding, equalization and synchronization, and localization. Likewise, DL capabilities have been studied for network resource optimization, proactive caching, routing and dynamic reconfiguration of antenna tilts for coverage optimization. However, the question remains if the DL-driven approaches are the right match to cope with the stringent requirements posed by highly dynamic SRE.
\par Recent studies~~\cite{zappone2019wireless}  suggest the use of DL approaches to support the development of novel methods enabling various RIS applications, as shown in Fig.~\ref{fig:App}, exploiting the spatially and temporally varying SRE. Notably, cutting edge DL techniques exhibit a strong potential of completely replacing conventional coding and modulation schemes, enabling pro-active and intelligent adaptation to the environment \cite{zappone2019wireless}. Likewise, for relay and WPT applications, feed-forward neural network has shown promise for modeling lossless propagation models, such that each RIS tile can be tuned to split and redirect its impinging power at its line-of-sight elements \cite{liaskos2019interpretable}. The input parameters include densities, location information of transmitters and receivers, noise levels, and dimensions of metasurface tiles; the activation function on the other hand is modeled as a combination of received signal strength and angle of arrival. 
\par Various DL solutions have been employed to learn the optimal re-configuration of metasurfaces once the radio waves impinge upon them, given that prior information about the propagation environment is made available. Taking into account the complex nature of interactions in RISs, especially for indoor environments, DL-based approaches exhibit their merit for signal focusing via learning the mapping between user position and the corresponding optimal configuration of metasurfaces~\cite{Huang}. Moreover, to support RIS-enabled multi-user communications, as shown in Fig.~\ref{fig:App}, emerging studies~\cite{zappone2019wireless} highlight the use of generative adversarial network (GAN) for channel agnostic learning; application of deep RL for solving the decentralized resource allocation problem; DL-assisted estimation of channel quality in mmWave massive MIMO systems, etc.; all indicate that the DL-assisted intelligence could be a major driver in realizing functionalities envisaged by ML-enabled RIS.
SRE, however, is fundamentally different from the existing networks and demands high level of network agility and adaptability. 
\par \textbf{\textit{Challenges:}}
The performance of DL models is limited by the amount of available training data and requisite computational power. The prohibitive training overhead, lack of efficient mechanisms to deal with parameter optimization of deep layered networks such that the models optimally converge within the coherence time of the wireless environment, and inability of existing approaches to adapt to heterogeneous hardware constraints for seamless connectivity as envisaged for RISs are few of the major challenges that demand resolution. 
\par  The preceding discussion on ML potentials and challenges is summarized in Fig.~\ref{fig:AI}. For the remainder of this section, we discuss further network design challenges and present  possible solutions to each challenge. 
\subsubsection{Network Agility}	
The reconfiguration of the metasurfaces relies on the amount of sensed data relayed back to the overarching network controller which is then processed for delay critical applications such as optimal beamforming. The aforementioned discussed technologies, such as deep neural networks could leverage the reported measurements to provide end-to-end intelligence and enable closed-loop network optimization. However, there exists a major concern in reducing the training time of these neural network algorithms, since the utility functions to be optimized are often complex and the number of parameters involved are often large, making the process computationally intensive. Conventional iterative methods utilized for training of these models induce latency, making them nonviable for RISs. However, there is an emerging research on exploiting the massive parallelism in the deep layered architecture of these models so that simultaneous weight calculation operations can be performed to reduce time and complexity of the learning process. Recent advances in \textit{graphical processing units}~\cite{zappone2019wireless} have also opened possibilities to speed up computation of these models (see Table~\ref{hard} for related hardware features). It has been widely argued that in order to make DL compatible with future wireless networks, distributed or on-device learning strategies must be leveraged. \textit{Federated learning techniques} can be employed, which work on a principle of distributing data and computational tasks among a federation of local resources governed by a central server. This approach could help devise a communication-efficient distributed training strategy for DL algorithms, since each resource processes data locally to learn a local DL model, whereas a central server learns a global model by integrating the distributed local models. To speed up the global model aggregation and reduce the communication overhead among the resources, it has been proposed to share the updated parameters only to a central server as opposed to the complete model. The distributed computing environment for both training and inference of DL-driven solutions could help design ultra-fast, low-power and low-cost processes critical for the emerging RISs network architecture. As detailed in Table~\ref{hard}, this can be achieved through the implementation of a nanonetwork.
\par To avoid the cost of training overhead, \textit{compressed sensing} tools have been used in order to construct  channels just by analyzing few of channel samples at  active elements \cite{Taha1}. 
\par \textit{Cross-fertilization} between model- and data-driven approaches is also an interesting research direction to explore in order to reduce the complexity and training overhead of the DL solutions. However, in a complex and dynamic environment surrounding RISs, the concept drift will be a frequent phenomenon. Therefore, the following question still remains: "knowing that the initial states, obtained from static models, may or may not be valid anymore, would the \textit{cross-fertilization} be able to yield optimized performances under such circumstances?" 

\subsubsection{Network Adaptability} 
SREs are anticipated to witness randomly evolving environments, as well as heterogeneous service requests, comprising different types of transceiver architectures (1-bit digital beamforming, analogue beamforming, etc.) and receivers. An outstanding issue is how to deal with the heterogeneity of RIS environments and effectively adapt to it without added complexity. \textit{Transfer learning} is an emerging technique that can help address this challenge by enabling the transfer of knowledge applied in a given context, to be used in a different context for executing a different task \cite{zhang2019deep}. One obvious advantage of using this technique is the reduction of training data, which has shown a great promise when particularly combined with DL techniques, being referred to as deep transfer learning (DTL). 

\par DTL methods have been broadly categorized into instance, network, adversarial, and mapping-based, and their applicability in wireless domains have been discussed in~\cite{zappone2019wireless}. The rationale behind using this approach is to transfer knowledge from a related reference scenario, where data acquisition has already been performed, to the target scenario -- mainly to avoid the expensive and time-consuming data acquisition process. For example, network-based DTL methods have been applied to identify the optimal deployment density of the base stations, and achieved a near-optimal performance comparing to data-driven models. Considering the SRE, where the sensed data is large and the environment is highly dynamic, it is not feasible to rely only on model-based or data-driven optimization techniques. Instead, the software controller ML-empowered RISs could exploit DTL techniques to adaptively optimize the system in real-time for evolving scenarios, by synergistically leveraging data- and model-driven approaches.

\section{A Look Ahead}\label{lookahead}
\subsection{Distributed Computation and Machine Learning}
While reconfigurable metasurfaces offer a promising platform for distributed computing and processing, the conventional distributed optimization methods are not well suited to guarantee real-time QoS needs of highly dynamic SREs. The complexity of RIS environments originates from the fact that not only there is large number of parameters that need to be optimized, but also the sensed data required for the optimized operation of smart radios is quite large and demands energy-efficient and time-critical processing. As discussed earlier in Section~\ref{cll}, metasurface-based modulation~\cite{Shan} is a promising research direction that could alleviate the challenge of data reduction and energy efficiency, since it modulates the sensed data onto the reflections or radiating patterns of nearby distributed devices; this, in turn, reduces the exchange of messages without additional energy cost. 
\par In principle, distributed ML techniques such as DL-based RL methods, discussed in Section~\ref{dnn}, offer promising solutions to enable dynamic network decision-making and to meet the stringent requirements of low-latency and real-time data processing. For example, wave transformations and modifications based on the subsequent response from a distributed metasurface environment requires learning, optimization and decision-making in a distributed manner. However, there are some issues that deserve attention: (a) how to reduce the communication overhead for scaling up the distributed learning and inference; (b) how to address the stringent computation, power, privacy, storage and bandwidth constraints for designing ultra-low-power, low-cost, and low-latency distributed inference mechanism. The merit of DTL techniques, which includes minimizing the communication overhead and embedding of expert knowledge into the ML models for faster inference, still needs to be explored. Federated learning techniques in conjunction with over-the-air-computation can be exploited to tackle issues like limited bandwidth, security and data privacy, since they enable learning of a shared global model from the local models computed individually by each distributed device. To meet the requirements of latency- and energy-aware communication of future networks, computing resources across the network including end distributed devices, network edge, and cloud must be leveraged simultaneously. 
\subsection{Quantum Machine Learning (QML)}
The emerging QML paradigm is receiving significant attention, since it is showing great promise for various applications in current and emerging communication networks, like RISs~\cite{oneto2017quantum}. QML attempts to remodel the ML problems either completely or in a hybrid fashion, such that parallelism offered by quantum computing can be exploited to accelerate the data processing speeds. A number of QML algorithms rely on the idea of amplitude information encoding, which is compact in nature and exponentially speeds up the matrix operations on vectors in high-dimensional vector spaces when compared to the classical counterparts. The ability to process information in a low-latency fashion is a critical metric for enabling real-time adaptivity of RIS functionalities. QML not only offers benefits of faster training and inference speeds in comparison to conventional ML approaches, but also improved security and privacy. A complexity comparison of state-of-the-art ML approaches and QML is provided in~\cite{mquantum}. Unlike conventional RL algorithms, which suffer from slow convergence time, quantum-powered RL leverages the superposition and parallelism concepts of quantum mechanics to speed up the learning, as discussed in~\cite{oneto2017quantum}. However, practical difficulties still exist. The development of quantum processors for quantum information processing is still underway. Moreover, scarcity of appropriate data sets, and also preparing an adequate input to be processed by quantum devices and subsequently extracting an output from them, are some of the unresolved issues. Notably, the crossover between quantum and deep neural networks is an interesting area which researchers have started to explore, and has remarkable scope for unlocking the full potential of RISs. 
\section{Conclusions}\label{conc}
As one of the promising technological paradigms envisioned for future 6G wireless networks, this paper presented an overview of RISs which can be programmed using software-defined functions. The discussed fundamental applications reveal that there exist non-trivial practical limitations, especially in future deployments, where the environment is expected to be highly dense and dynamic. Therefore, as ML mechanisms are foreseen to be inherent in the optimization of RISs-enabled wireless networks, we offered a detailed discussion on the potentials and limitations of major ML approaches studied in the area of RISs.

\bibliographystyle{IEEEtran}
\bstctlcite{BSTcontrol}
\bibliography{Reflist}

\begin{thebibliography}{10}
\providecommand{\url}[1]{#1}
\csname url@samestyle\endcsname
\providecommand{\newblock}{\relax}
\providecommand{\bibinfo}[2]{#2}
\providecommand{\BIBentrySTDinterwordspacing}{\spaceskip=0pt\relax}
\providecommand{\BIBentryALTinterwordstretchfactor}{4}
\providecommand{\BIBentryALTinterwordspacing}{\spaceskip=\fontdimen2\font plus
\BIBentryALTinterwordstretchfactor\fontdimen3\font minus
  \fontdimen4\font\relax}
\providecommand{\BIBforeignlanguage}[2]{{%
\expandafter\ifx\csname l@#1\endcsname\relax
\typeout{** WARNING: IEEEtran.bst: No hyphenation pattern has been}%
\typeout{** loaded for the language `#1'. Using the pattern for}%
\typeout{** the default language instead.}%
\else
\language=\csname l@#1\endcsname
\fi
#2}}
\providecommand{\BIBdecl}{\relax}
\BIBdecl

\bibitem{zappone2019wireless}
A.~{Zappone}, M.~{Di Renzo}, and M.~{Debbah}, ``Wireless networks design in the
  era of deep learning: Model-based, {AI}-based, or both?'' \emph{IEEE Trans.
  Commun.}, vol.~67, no.~10, pp. 7331--7376, Oct. 2019.

\bibitem{Liaskos1}
C.~{Liaskos}, S.~{Nie}, A.~{Tsioliaridou}, A.~{Pitsillides}, S.~{Ioannidis},
  and I.~{Akyildiz}, ``A new wireless communication paradigm through
  software-controlled metasurfaces,'' \emph{IEEE Commun. Mag.}, vol.~56, no.~9,
  pp. 162--169, Sept. 2018.

\bibitem{Huang}
C.~Huang, G.~C. Alexandropoulos, C.~Yuen, and M.~Debbah, ``Indoor signal
  focusing with deep learning designed reconfigurable intelligent surfaces,''
  in \emph{Proc. IEEE Int. Workshop . Signal Process. Advances in Wireless
  Commun.}\hskip 1em plus 0.5em minus 0.4em\relax Cannes, France, July 2019,
  pp. 1--5.

\bibitem{9007666}
S.~{Abadal}, T.~{Cui}, T.~{Low}, and J.~{Georgiou}, ``Programmable
  metamaterials for software-defined electromagnetic control: Circuits,
  systems, and architectures,'' \emph{IEEE Trans. Emerg. Sel. Topics Circuits
  Syst.}, vol.~10, no.~1, pp. 6--19, Mar. 2020.

\bibitem{Wang}
J.~{Wang} \emph{et~al.}, ``Metantenna: When metasurface meets antenna again,''
  \emph{IEEE Trans. Antennas Propag.}, vol.~68, no.~3, pp. 1332--1347, Mar.
  2020.

\bibitem{8741198}
C.~{Huang}, A.~{Zappone}, G.~C. {Alexandropoulos}, M.~{Debbah}, and C.~{Yuen},
  ``Reconfigurable intelligent surfaces for energy efficiency in wireless
  communication,'' \emph{IEEE Trans. Wireless Commun.}, vol.~18, no.~8, pp.
  4157--4170, Aug. 2019.

\bibitem{Shan}
T.~{Shan}, X.~{Pan}, M.~{Li}, S.~{Xu}, and F.~{Yang}, ``Coding programmable
  metasurfaces based on deep learning techniques,'' \emph{IEEE Trans. Emerg.
  Sel. Topics Circuits Syst.}, vol.~10, no.~1, pp. 114--125, Mar. 2020.

\bibitem{ntontin2019reconfigurable}
\BIBentryALTinterwordspacing
K.~Ntontin \emph{et~al.}, ``Reconfigurable intelligent surfaces vs. relaying:
  Differences, similarities, and performance comparison,'' \emph{arXiv preprint
  arXiv:1908.08747}, 2019. [Online]. Available:
  \url{https://arxiv.org/abs/1908.08747}
\BIBentrySTDinterwordspacing

\bibitem{MS}
C.~Liaskos, S.~Nie, A.~Tsioliaridou, A.~Pitsillides, S.~Ioannidis, and
  I.~Akyildiz, ``A novel communication paradigm for high capacity and security
  via programmable indoor wireless environments in next generation wireless
  systems,'' \emph{Ad Hoc Networks}, vol.~87, pp. 1 -- 16, May 2019.

\bibitem{simeone2018very}
O.~Simeone, ``A very brief introduction to machine learning with applications
  to communication systems,'' \emph{IEEE Trans. Cognitive Commun. Netw.},
  vol.~4, no.~4, pp. 648--664, Dec. 2018.

\bibitem{liaskos2019interpretable}
C.~{Liaskos}, A.~{Tsioliaridou}, S.~{Nie}, A.~{Pitsillides}, S.~{Ioannidis},
  and I.~{Akyildiz}, ``An interpretable neural network for configuring
  programmable wireless environments,'' in \emph{IEEE Int. Workshop Signal
  Process. Advances in Wireless Commun. ({SPAWC}'19)}, July 2019, pp. 1--5.

\bibitem{zhang2019deep}
C.~Zhang, P.~Patras, and H.~Haddadi, ``Deep learning in mobile and wireless
  networking: A survey,'' \emph{IEEE Commun. Surveys Tuts.}, vol.~21, no.~3,
  pp. 2224--2287, Third Quarter 2019.

\bibitem{Taha1}
\BIBentryALTinterwordspacing
A.~Taha, M.~Alrabeiah, and A.~Alkhateeb, ``Enabling large intelligent surfaces
  with compressive sensing and deep learning,'' \emph{arXiv:1904.10136v2},
  2019. [Online]. Available: \url{https://arxiv.org/abs/1904.10136}
\BIBentrySTDinterwordspacing

\bibitem{oneto2017quantum}
L.~Oneto, S.~Ridella, and D.~Anguita, ``Quantum computing and supervised
  machine learning: Training, model selection, and error estimation,'' in
  \emph{Quantum Inspired Computational Intelligence}.\hskip 1em plus 0.5em
  minus 0.4em\relax Elsevier, Chapter 2 2017, pp. 33--83.

\bibitem{mquantum}
C.~Outeiral, M.~Strahm, J.~Shi, G.~M. Morris, S.~C. Benjamin, and C.~M. Deane,
  ``The prospects of quantum computing in computational molecular biology,''
  \emph{WIREs Computational Molecular Science}, p. e1481.
\renewcommand{\BIBentryALTinterwordstretchfactor}{4}

\end{thebibliography}

\begin{IEEEbiographynophoto}
{Lina Mohjazi} (M’18, SM'20) is a Lecturer at the School of Engineering, University of Glasgow, UK. She received the M.Sc. degree from Khalifa University, UAE, in 2012, and the Ph.D. degree from the University of Surrey, UK, in 2018, both in electrical and electronic engineering. She is currently a Review Editor for Frontiers in Communications and Networks and an Area Editor for Physical Communication (Elsevier). Her main research interests include beyond 5G wireless technologies, physical layer optimization and performance analysis, wireless power transfer, machine learning for future wireless systems, and reconfigurable intelligent surfaces.

\end{IEEEbiographynophoto}
\vspace{-1cm}
\begin{IEEEbiographynophoto}
{Ahmed Zoha} is a lecturer at the School of Engineering, University of Glasgow, UK. He received his Ph.D. from the Institute of Communication Systems (ICS) at the University of Surrey and M.Sc. in Communication Engineering from Chalmers University of Technology, Sweden. Dr. Ahmed has more than 10 years of experience in the domain of quantitative analysis and research, data mining, predictive modeling, and applied machine learning and artificial intelligence for real-world applications. Dr. Ahmed has contributed to several interdisciplinary and multimillion-funded international research projects in the domain of big data enabled self-organizing wireless communication networks, smart energy monitoring and connected healthcare.  His research interests include machine learning at the edge, 5G and beyond, digital health and internet of everything. (e-mail: ahmed.zoha@glasgow.ac.uk) 
\end{IEEEbiographynophoto}
\vspace{-1cm}
\begin{IEEEbiographynophoto}
{Lina Bariah} (S’13, M’19) received the M.Sc. and Ph.D degrees in communications engineering from Khalifa University, Abu Dhabi, United Arab Emirates, in 2015 and 2018. She is currently a Postdoctoral fellow with the KU Center for Cyber-Physical Systems, Khalifa University, UAE. Her research interests include advanced digital signal processing techniques for communications, channel estimation, cooperative communications, non-orthogonal multiple access and cognitive radios.
\end{IEEEbiographynophoto}
\vspace{-1cm}
\begin{IEEEbiographynophoto}
{Sami Muhaidat} (M'07, SM'11)  received the Ph.D. degree in electrical and computer engineering from the University of Waterloo,  Ontario, in 2006. From 2008-2012 he was an Assistant Professor in the School of Engineering Science, Simon Fraser University, Canada. He is currently a Full Professor at Khalifa University. His research focuses on physical layer aspects of wireless communications. He is currently an Area Editor for IEEE Transactions on Communications. Previously, he served as a Senior Editor for IEEE Communications Letters.
\end{IEEEbiographynophoto}
\vspace{-1cm}
\begin{IEEEbiographynophoto}
{Paschalis C. Sofotasios} (M’12, SM’16) received the M.Eng. degree from Newcastle University, U.K., in 2004, the M.Sc. degree from the University of Surrey, U.K., in 2006, and the Ph.D. degree from the University of Leeds, U.K., in 2011. He has held academic positions at the University of Leeds, U.K., and the Khalifa University, UAE, where he currently serves as an Assistant Professor. His research interests include broad areas of digital and optical wireless communications. He served as an Editor for the IEEE Communications Letters.
\end{IEEEbiographynophoto}
\vspace{-1cm}
\begin{IEEEbiographynophoto}
{Muhammad Ali Imran} (M'03, SM'12) Fellow IET, Senior Member IEEE, Senior Fellow HEA is a Professor of Wireless Communication Systems with research interests in self organized networks, wireless networked control systems and the wireless sensor systems. He heads the Communications, Sensing and Imaging CSI research group at University of Glasgow and is the Dean University of Glasgow, UESTC. He is an Affiliate Professor at the University of Oklahoma, USA and a visiting Professor at 5G Innovation Centre, University of Surrey, UK. He has over 20 years of combined academic and industry experience with several leading roles in multi-million pounds funded projects. He has filed 15 patents; has authored/co-authored over 400 journal and conference publications; was editor of 5 books and author of more than 20 book chapters; has successfully supervised over 40 postgraduate students at Doctoral level. He has been a consultant to international projects and local companies in the area of self-organised networks. He has been interviewed by BBC, Scottish television and many radio channels on the topic of 5G technology.
\end{IEEEbiographynophoto}
\vspace{-1cm}
\begin{IEEEbiographynophoto}
{Octavia A. Dobre} (M'05, SM'07, F'20) is a Professor and Research Chair at Memorial University (Canada), which she joined in 2005. Octavia received the Dipl. Ing. and Ph.D. degrees from the Polytechnic University of Bucharest (formerly Polytechnic Institute of Bucharest), Romania in 1991 and 2000, respectively. She was a Royal Society Scholar, and a Fulbright Scholar. She is a fellow of the Engineering Institute of Canada and a Distinguish Lecturer of the IEEE Communications Society. She is the recipient of diverse awards, including Best Paper Awards at IEEE ICC and IEEE WCNC. Her main research areas are NOMA, full-duplex, signal identification, as well as optical and underwater communication. 
\end{IEEEbiographynophoto}
\end{document}